# Unveiling the Interstitial Pressure Between Growing Ice Crystals During Ice-Templating Using a Lipid Lamellar Probe

**Niki Baccile,[a,*] Thomas Zinn,[b] Guillaume P. Laurent,[a] Ghazi Ben Messaoud,[a,†] Viviana Cristiglio,[c] Francisco M. Fernandes[a,*]**

[a] Sorbonne Université, Centre National de la Recherche Scientifique, Laboratoire de Chimie de la Matière Condensée de Paris, LCMCP, F-75005 Paris, France

[†] Current address: DWI- Leibniz Institute for Interactive Materials, Forckenbeckstrasse 50, 52056 Aachen, Germany

[b] ESRF - The European Synchrotron, 71 Avenue des Martyrs, 38043 Grenoble, France

[c] Institut Laue-Langevin, 71 Avenue des Martyrs, 38042 Grenoble Cedex 9, France

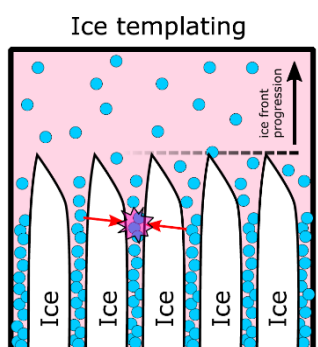

**Abstract** What is pressure generated by ice crystals during ice-templating? This work addresses this crucial question by estimating the pressure exerted by oriented ice columns on a supramolecular probe composed of a lipid lamellar hydrogel during directional freezing. This process, also known as freeze-casting, has emerged as a unique processing technique for a broad class of organic, inorganic, soft and biological materials. Nonetheless, the pressure exerted during and after crystallization between two ice columns is not known, despite its importance with respect to the fragility of the frozen material, especially for biological samples. By using the lamellar period of a glycolipid lamellar hydrogel as a common probe, we couple data obtained from ice-templated-resolved *in situ* synchrotron Small Angle X-ray Scattering (SAXS) with data obtained from controlled adiabatic dessication experiments. We estimate the pressure to vary between $1 \pm 10\%$ kbar at -15°C and $3.5 \pm 20\%$ kbar at -60°C.

**Introduction** Freezing is a universal strategy for long-term preservation of complex aqueous systems. From food conservation to cell banking, it remains the simplest – yet most effective – approach to minimize water activity, while maintaining the molecular and/or biological integrity of the systems to preserve. In most cases, freezing aqueous systems result in phase segregation between pure ice and a solute-rich phase confined within the interstitial space defined by ice crystals.[1] It is there, in the micrometer range, in between growing ice crystals, that major transformations occur, imposing dramatic variations to conditions such as hydration, temperature, pressure, stress and strain. These processes have huge implications in fields spanning from geology,[2] cultural heritage[3] and civil engineering,[4] materials science[5] and chemistry,[6] with a particularly strong impact in life science.[7] In the latter, the course of the out-of-equilibrium freezing phenomena defines the environmental conditions endured by the entities to be frozen and, thus, it dictates the outcome of conservation. Among these conditions, pressure plays a central, yet still unclear, role. The stability of biomolecules to pressure has been mostly conducted under hydrostatic conditions. It has been shown that nucleic acids are able to withstand pressures up to 12 kbar without enduring any sensible denaturation process,[8] while proteins unfold between 1.5 and 6 kbar.[9,10] Microorganisms, which are not particularly adapted to extreme pressures like *E. coli*, retain some biological function up to 3 kbar,[11] whereas the critical range for eukaryotic cells lies between 1 kbar and 3 kbar.[12]

Despite conceptually simple, measuring the local pressure in between growing ice crystals is far from being trivial: it implies to precisely control the growth of ice throughout the experiment and, above all, to use a relevant probe to sense the generated pressure. Previous attempts at unveiling the pressures involved in water freezing have resulted in scattered values ranging from 0.2 bar in a simple air cooled setup, where pressure was followed by an optical microscope,[13] to 200 bar in lipid lamellar phases plunged into liquid nitrogen and measured by $^2$H NMR spectroscopy[14]. The lack of congruence found in the values reported from such experiments results from the differences in experimental design, namely the geometrical constraints imposed on the frozen materials, the thermal boundary conditions, as well as in the technical limits to discriminate between nucleation and growth steps during ice crystallization.

To account for these limitations, we have devised a new strategy based on ice-templating, where both geometrical and thermal boundaries can be precisely defined. Ice-templating – also referred to as freeze-casting – is a directional freezing technique that has emerged as a materials processing technique suitable for the elaboration of porous ceramics.[5] Since then, it has evolved as a straightforward technique to cast a myriad of materials from graphene oxides[15] to polysaccharides,[16,17] proteins[18] and physical hydrogels.[19] To locally measure the pressure in the interstitial space formed between ice crystals, we have selected a lipid lamellar hydrogel that has been previously shown to withstand the freezing process without loss of structure.[19] We monitor the interlamellar distance of the lipid gel by coupling a dedicated ice-templating setup to temperature-resolved small angle X-ray scattering (SAXS) in order to achieve an *in situ* time- and space-resolved relationship between the ice front progression and the interlamellar space during freezing. To infer the pressure from the SAXS experiments, we have measured the interlamellar distance variation, on the same lamellar gel, due to osmotic stress under adiabatic conditions, at room temperature, with neutron diffraction. In both cases, the pressure rise is a consequence of gel dehydration, imposed by water freezing, in one case, and by decrease in relative humidity, in the other. By combining both experiments, dehydration by directional freezing and adiabadic dessication,[20] we depict a broad set of pressure-temperature relationships corresponding to a variety of freezing conditions, where interstitial pressure generated by the ice crystals can be estimated to lie between 1 kbar and 5 kbar, a critical range in cryobiology.

These results enable to better understand a critical aspect in the design of future cryopreservation systems by discriminating between local pressures generated in two clearly distinct zones respectively dominated by ice nucleation and growth. The understanding of these crystallization regimes directly impacts a diversity of fields spanning from cryobiology to material science, chemistry, biochemistry, thus providing elements to rationalize pressure effects related to freezing.

**Methodology**

The pressure-temperature relationship in the ice-templating experiment has been determined according to the methodology illustrated in Figure 1. The GC18:0 lamellar hydrogel (C = 10 wt%, pH = 6.2, [NaCl] = 50 mM) is employed as probe to indirectly measure the pressure generated by directional water freezing in an ice-templating device. *In situ* SAXS (Figure S 1) data acquisition during freezing enables to measure the evolution of the lamellar period as a function of temperature, $d_{(100)}(T)$ (Figure S 2, Figure S 3). $d_{(100)}(T)$ is used as the actual probe to evaluate the pressure exerted by ice under unidirectional freezing and controlled conditions of cooling rate (5°C·min$^{-1}$ or 10°C·min$^{-1}$). The temperature gradient imposed by the ice-templating setup induces ice growth along the *Z* axis (Figure 1). Before ice is formed, the lamellar domains are mainly contained in the *XY* plane (T > -10°C).[19] Throughout the freezing process, the lamellar domains tilt to the *XZ* plane (T < -10°C, director of lamellar phase parallel to *Y* axis, Figure 1) to accommodate for the ice columns growth.[19] Importantly, the orientation imposed by the ice columns in the growth on the lamellar domains (lamellar director parallel to the *Y* axis) corresponds to the expected pressure direction in between ice crystals.

To transform $d_{(100)}(T)$ into a pressure-temperature relationship, we study the pressure-distance relationship, $\Pi(d_{(100)})$, of the same GC18:0 lamellar hydrogel using the isothermal (T = 25°C) osmotic stress technique[21,22] inside an adiabatic humidity chamber (Figure S 4a).[23,24] The hydrogel is drop-cast and allowed to dry on a silicon wafer, having the lamellar director parallel to the *Y* axis, normal to the silicon wafer. The lamellar spacing is probed using neutron diffraction in a $\theta$-$2\theta$ configuration, with the relative humidity (*RH%*) varying between 98% and 10% (Figure S 4b).[25] The distance-humidity relationship, $d_{(100)}(RH\%)$, (Figure S 4b) is converted into a $\Pi(d_{(100)})$ relationship, using the following expression that equalizes pressure and *RH%*,[26,27]

Eq. 1
$$\Pi = -N_A \left(\frac{k_b T}{V_w}\right) ln\left(\frac{RH\%}{100}\right)$$

with $\Pi$ being the osmotic pressure, $N_A$ the Avogadro constant, $k_b$ the Boltzmann's constant, $T$ the temperature in degrees Kelvin, $V_w$ the water molar volume and *RH%* the relative humidity. Under these conditions, $d_{(100)}$ identifies the lamellar period and it constitutes the common parameter, experimentally measured on the same material, between the humidity chamber and the ice-templating device. The analogy between these systems is based on the residual hydration of the lamellar phase at temperatures as low as -60°C and probed by solid state $^2$H NMR experiments performed from +20°C to -60°C using a cooling rate of 10°C·min$^{-1}$ (Figure S 5).

Finally, when the membrane thickness is subtracted from $d_{(100)}$, one obtains the thickness of the interlamellar water layer, $d_w$, which is assumed to be invariable throughout the process.

The pressure-distance relationship, $\Pi(d_w)$, well-known in membrane physics,[22,28,29] allows to associate the intermembrane osmotic pressure, $\Pi$, to $d_w$, by calculating each term of the equation of state (Eq. 2) composed of attractive (Van der Waals, $\Pi_{VdW}$) and repulsive forces (hydration, steric, electrostatic, respectively, $\Pi_{Hyd}$, $\Pi_{St}$, $\Pi_{El}$). In ref.[25], we study the equation of state of the GC18:0 lamellar hydrogel at $d_w <$ 3 nm and we show that the repulsive term of $\Pi(d_w)$ is only composed of the short-range ($d_w <$ ~0.5 nm) hydration force, $\Pi_{Hyd1}$. However, in the same work we could also show that the presence of salt introduces an additional hydration term, $\Pi_{Hyd2}$, in the equation of state, well-known in the literature for condensed lamellar phases at high ionic strength, commonly referred as secondary hydration and acting at distances between 0.5 nm and 3 nm.[30–32] Eq. 2 can be then reduced to Eq. 3 for $d_w <$ 3 nm and of which the extended expressions and numerical values are given in the Supporting information: Eq. S 6, Eq. S 7 and Table S1 for the hydration terms and Eq. S 5 and section SI 5 for the Van der Waals term. Introducing $d_w(T)$, measured by controlled directional freezing of the GC18:0 lamellar gel, in the $\Pi(d_w)$ relationship established by the osmotic stress technique on the same material, we establish the $\Pi(T)$ relationship associated to ice-templating (Eq. 4). The resulting data are summarized in Figure 2 and Figure S 6.

$$\Pi = \Pi_{VdW} + \Pi_{St} + \Pi_{Hyd} + \Pi_{El} \qquad \text{Eq. 2}$$

$$\Pi(d_w) = \Pi_{VdW}(d_w) + \Pi_{Hyd1}(d_w) + \Pi_{Hyd2}(d_w) \qquad \text{Eq. 3}$$

$$\Pi(T) = \Pi_{VdW}(T) + \Pi_{Hyd1}(T) + \Pi_{Hyd2}(T) \qquad \text{Eq. 4}$$

Sections SI 3 and SI 5 in the Supporting Information summarize the methodology employed in this work concerning the calibration pressure-distance curves presented in ref.[25]. Before discussing the data, we address some important points, that validate the present approach.

(i) *The validity of the lipid lamellar phase as a pertinent model*. Lipid lamellar phases are long-studied systems governed by attractive (Van der Waals) and repulsive (steric, hydration, electrostatic) forces, generally described at thermodynamic equilibrium within the framework of an extended DLVO theory, including hydration forces and thermal fluctuations.[22,27,28,33,34] The most interesting feature of lipid lamellar phases is the reversibility of the interlamellar distance variation due to external osmotic stress.[27,35,36] The corresponding pressures range between fractions of bar to several kbar,[27,35,37,38] and generally without affecting the bilayer structure. To this regard, a lipid lamellar phase is an interesting soft material to study the interstitial pressure associated to the directional growth of ice columns (Figure 1).

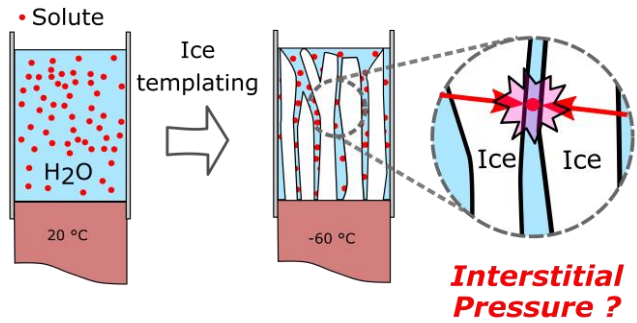
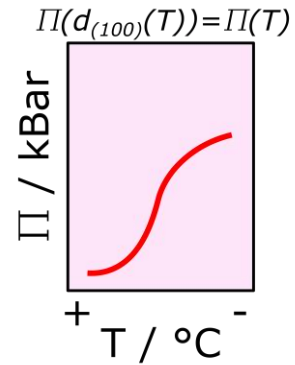
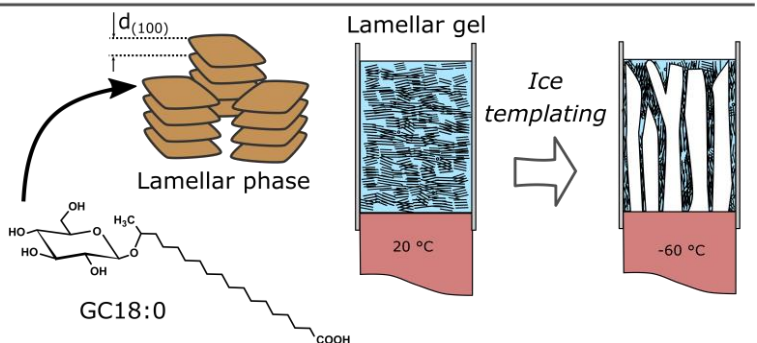
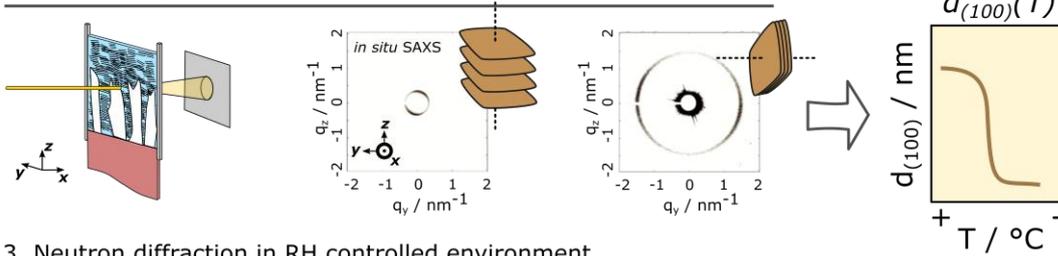
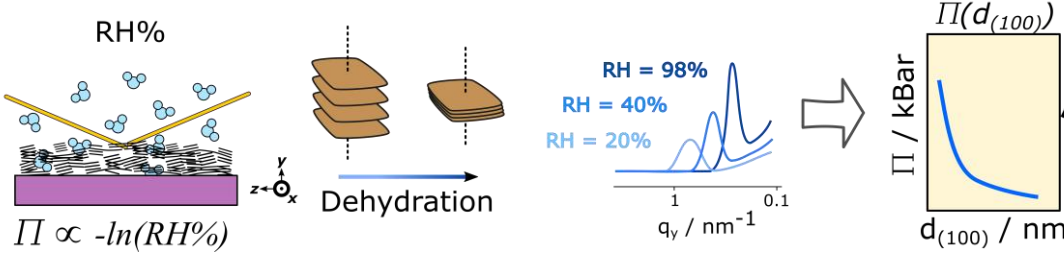

**Figure 1: Which is the interstitial pressure exerted by ice on a solute during directional freezing? We address this problem by: 1. employing a lipid lamellar probe; 2. measuring the evolution of the lamellar period with temperature, $d_{(100)}(T)$, by coupling *in situ* synchrotron SAXS to an ice-templating device; 3. measuring the evolution of the lamellar period with osmotic pressure, $\Pi(d_{(100)})$, under controlled relative humidity, RH%, at 25°C under adiabatic conditions. Finally, $\Pi(d_{(100)})$ and $d_{(100)}(T)$ are combined to obtain $\Pi\left(d_{(100)}(T)\right) = \Pi(T)$, describing the osmotic pressure associated to a given temperature during ice-templating. The pressure is always exerted orthogonal to the lamellar (Y-axis) plane both during ice-templating and dehydration experiments. Hydration of the lamellar phase during ice-templating is probed**

**by $^2$H solid state NMR between +20°C and -60°C both under kinetic (10°C·min$^{-1}$ ) and thermodynamic conditions.**

(ii) *The compatibility between the conditions of ice-templating with isothermal osmotic stress measurements.* The pertinence in associating osmotic stress experiments performed at room temperature to freezing of lamellar phases has long been addressed, because both processes are simply governed by dehydration of the interlamellar water phase.[20] In both cases, the effect of hydration/dehydration on the pressure-distance curves in lipid lamellar phases has been studied for several years, and residual hydration generating strong repulsive forces (either *RH%* < 30% or T < -20°C) between the lamellae is the common denominator in both experiments.[14,20,27,29,39–43] The physicochemical analogy between the osmotic stress and freezing experiments was shown to hold for phospholipid lamellar phases, the most studied systems in the literature.[14,20,40,41] Typical bias in associating these experiments can be interlamellar ice nucleation or generation of strong intramembrane stresses due to extreme drying and modifying the membrane mechanical properties. These aspects, among others, are evidenced in ref. 20, which shows, along with other studies,[14,41,43] that ice nucleation is favored outside rather than within the lipid intralamellar space, generating dehydration. Solute (e.g. salts) concentration increases in the intralamellar space, which becomes an even less favorable site for ice nucleation.

(iii) *Impact of solutes in the validity of the model*. The presence of salt ions alongside with the lipid molecules is instrumental in the formation of the lamellar gel system.[44] However, presence of solutes adds supplementary complexity to the pressure-temperature relationship, both at room temperature[38,45] and below the freezing point of water.[14] Owing to the limited solubility of most solutes in ice (hexagonal ice is known to form a limited number of solid solutions with few compounds),[20,46,47] crystallization of solutes during freezing, and of course during dehydration, are known facts. One could then argue that dehydration of the solute-rich phase due to ice formation progressively concentrates solutes, especially in confined systems.[48,49] It would not then be unreasonable that such solute pouches could generate an extra pressure on the lipid phase. However, it was shown that these arguments do not apply to lipid/solute/water systems neither at room temperature[38,45] nor below the freezing point of water.[14,50] Small solutes (ions, sugars) are known to be intimately associated to lipid membranes,[45,51] especially at moderate/small concentrations below the molar range, which is the regime explored in this work. Even in dehydrated systems, at small distances (< 1 nm), small solutes concentrate in the interlamellar volume. This is a general feature of lamellar systems and it was shown to hold up to 5 M salt concentrations in mica systems.[32] Large solutes, like polymers (not used in this work), are, on the contrary, known to be expelled from the intermembrane space upon dehydration.[50]

(iv) *The temperature-dependency of the equation of state.* Several terms of the equation of state (Eq. 2) are known to depend on temperature. If $\Pi_{El}$ depends on temperature,[52] the short interlamellar distances studied here excludes this term.[25] An additional notoriously temperature-dependent term is the long-range entropic (undulation) contribution,[53] voluntarily excluded from this work.[25] The Van de Waals term, $\Pi_{VdW}$, also depends on temperature through the Hamaker constant. In the Supporting Information (Page 13, section SI 5), we address the problem of calculating the Hamaker constant below the freezing point of water. One finds a value close enough (< 15%) to the typical value at room temperature ($H = 5.1 \cdot 10^{-21}$ J) to exclude any substantial impact on our calculation of $\Pi_{VdW}$, when considering the global uncertainty of our approach. Finally, the hydration term of Eq. 2 is commonly accepted as being temperature-independent,[26,54,55] even below the freezing point of water.[14,56] In the few cases where the hydration term showed a dependency on temperature, the variation was generally reported to be less than 10%,[27,57] enough to be considered as not relevant within the framework of this work.[58]

(v) *The pertinence of the lamellar hydrogel employed here*. Concerning the choice of the sample employed as probe in this work, we have recently shown that the GC18:0 (Figure 1) molecule assembles into an interdigitated lipid $P_{\beta,i}$ lamellar phase forming a physical hydrogel in water at concentrations above 1 wt% and $T < 30°C$.[44] Under typical conditions (C = 5 wt% and 10 wt%, pH = 6.2 ± 0.3, [NaCl] = 50 mM), the hydrogel is composed of a lamellar phase with equilibrium lamellar period at room temperature of about $d_{(100)}$= 20 nm (Figure S 2a,b).[19] Hydrogels are convenient for their adaptability to both surface (osmotic stress in the humidity chamber) and bulk sample environments (ice-templating device). In addition, we have found that the variation in the lamellar period of GC18:0 hydrogels is fully reversible, both at room temperature, if pH or ionic strength are varied, and over a temperature cycle between +20°C and -60°C (Figure S 3b).[19,44] The melting temperature of GC18:0 is at about 37°C,[44] meaning that the interdigitated lipid layer is always in a solid-like gel phase below $T = 30°C$. This is an ideal condition, even at very low temperatures, because it avoids possible variations in thickness coming from fluid to gel transitions, generally induced by strong intralamellar stresses.[20]

(vi) *The possible temperature-dependency of the lipid membrane thickness.* SAXS experiments performed on both diluted[59,60] and concentrated[44] GC18:0 lamellar solutions at room temperature provide a thickness of the interdigitated layer of about 3.6 nm (± 10 %) in water at pH between 6 and 7 and at [NaCl] = 50 mM. It is know that heating above the lipid $T_m$ reduces the membrane thickness.[36] However, below the $T_m$ it is reasonable to assume that lipid bilayers are incompressible.[20] In the present system, the lipid is always studied below its $T_m$, thus excluding important variations in the thickness at temperature below room temperature. In fact, only tilting of the lipid membrane could have an impact on the thickness upon freezing. In the case of a reasonable 30° tilt angle, the error on the evaluation of the thickness would be less than 15%, that is within the experimental error of our method. Using SAXS data, we show in Figure S 3 and discuss in section SI 2 in the Supporting Information, that the broad oscillation at q > 0.5 nm$^{-1}$, characteristic of the bilayer form factor (which includes membrane thickness and electron density), remains unchanged both between room temperature and $T$= -8°C – just before ice crystallization – and at room temperature after two freezing cycles between $T$= +20°C and $T$= -60°C. The thickness is then considered as constant. In addition, during the evolution of the $d_{(100)}$-spacing during ice-templating from +20°C to -60°C, we find that the lowest reachable d-spacing is $d_{(100)}$ = 4.04 ± 0.01 nm (Figure S 2b).[44] To this value one should subtract at least a single hydration layer (0.28 nm, taken as the diameter of one H$_2$O molecule) and/or a single layer of counterion (0.33 nm for Cl$^-$), as discussed in ref. [20]. The estimated thickness of the GC18:0 interdigitated layer at -60°C is then in the order of 3.7 nm, in very good agreement, within the error, with the value measured in solution at room temperature,[60] used in this work.

(vii) *The nature of the interface between the lipid membrane and the water gas and solid phase*. It is important to prove that the nature of the interface between the lipid membrane and the water phase (gas or solid), which apply pressure, is the same. In the experiments performed at room temperature in the adiabatic humidity chamber, the interface is obviously constituted by a thin layer of liquid water. We have then run a series of $^2$H solid-state NMR experiments during freezing to prove the presence of liquid water below freezing. The concentration of solute in between the GC18:0 lamellae upon freezing should ensure interlamellar hydration, as expected for phospholipid-solute lamellar systems,[14,20,41,43] and confirmed by $^2$H solid-state NMR experiments performed in this work on the GC18:0 lamellar hydrogel itself (Figure S 5a,b). These experiments have been voluntarily performed using a freezing rate of 10°C·min$^{-1}$ between +20°C and -60°C (Figure S 5a) but also at equilibrium, in order to follow the hydration at low temperature under both kinetic, the closest experimental conditions to the ice-templating experiment, and thermodynamic conditions. The relative content of liquid-like water is estimated for each spectrum in Figure S 5a,b by fitting the entire $^2$H spectrum with a broad quadrupolar (solid) and narrow Lorentzian (liquid) component (e.g. in Figure S 5c). $^2$H NMR shows the presence of the central peak at δ= 0 ppm from -20°C to -60°C (Figure S 5b), confirming that a typical 5 wt% gel contains between 5 wt% and 1 wt% of liquid-like water (Figure S 5a), respectively corresponding to 25 and 5 water molecules per GC18:0 molecule, in such temperature range. Similar experimental conditions were used to analyze a control sample of pure D$_2$O to clarify the role of the lamellar phase in the stabilization of liquid water up to -60 °C (Figure S 5a). We find that water is qualitatively frozen below -10°C. The existence of liquid water in the lamellar hydrogel at temperature as low as -60°C, and compared to pure D$_2$O, comforts the analogy between freezing and dehydration, thus validating our analytical approach. An equivalent $^2$H solid-state NMR experiment has been conducted on both GC18:0 lamellar hydrogel and pure D$_2$O under thermal equilibrium conditions in the same temperature range to highlight differences that are inherent to the out-of-equilibrium nature of ice-templating (Table S 2): the lamellar hydrogel retains liquid water at temperatures as low as -40°C.

**Pressure-temperature, $\Pi(T)$, relationship generated in an ice-templating/freeze-casting process**

The evolution of $d_{(100)}$ as a function of the freezing temperature, $d_{(100)}(T)$, collected at various positions of the ice-templating device and for two freezing rates, is obtained from *in situ* SAXS experiments (Figure S 2) collected on a typical GC18:0 hydrogel (C = 10 wt%, pH 6.2 ± 0.3) prepared at [NaCl] = 50 mM. The corresponding interlamellar water layer thickness, $d_w$, is calculated $d_w(T) = d_{(100)}(T) - 3.6$ nm, and $d_w(T)$ values are inserted in $\Pi(d_w)$ (Eq. 3), so to obtain $\Pi(T)$ (Eq. 4) associated to the ice-templating process (Figure 2). $\Pi(d_w)$ profiles are empirically determined by osmotic stress experiments in the humidity chamber (Figure S 4a) collected on two GC18:0 hydrogel samples (concentration before drop-cast, C = 1 wt%, pH = 6.2 ± 0.3), prepared in a low-salt ([NaCl] = 16 mM) and high-salt ([NaCl] = 100 mM) regime. From these experiments, one can determine analytical expressions of $\Pi(d_w)$, which was done in Ref. [25] using four different fitting strategies. Section SI 5 in the Supporting Information presents the $\Pi(T)$ profiles associated to each ice-templated system. Each curve (red squares in Figure S 6) is averaged after the use of four different $\Pi(d_w)$ corresponding to the four fits,[25] which also determine an empirical confidence range (yellow regions in Figure S 6).

The general pressure-temperature, $\Pi(T)$, profiles associated to the ice-templating process using the lamellar GC18:0 gel as probe are shown in Figure 2. They are given for two freezing rates 5°C·min$^{-1}$ and 10°C·min$^{-1}$, respectively, corresponding to ice-front speed of 9.16 ± 0.59 µm/s and 15.44 ± 2.4 µm/s, measured in ref. [44], and at four different heights, $h$, in the device. Positions $h$ = 100 µm and $h$ = 1700 µm are respectively the closest and the farthest with respect to the cold metal surface (Figure S 1). The empty blue symbols show the raw $d_{(100)}(T)$ profiles from which $d_w(T)$ are calculated and employed in Eq. 3 to calculate $\Pi(T)$ (Eq. 4). The average values of $\Pi(T)$ and the corresponding confidence range determined with fits (1)-(4)[25] for low-ionic strength (intersecated grey symbols, Figure 2a,b-bottom) and high- ionic strength (half-filled grey symbols) regimes are respectively represented as filled color symbols and yellow region.

The domain of interstitial pressures generated between ice columns in the ice-templating apparatus are contained between 1 kbar and 5 kbar under all conditions explored. At both 5°C·min$^{-1}$ and 10°C·min$^{-1}$ and for distances above $h$ = 500 µm (Figure 2c,d), only the secondary long-range hydration regime is reached, characterized by the structuring of water around the counterions[38,61,62] and with a decay lengths of about 2 nm (Ref. [25] and Supporting Information, section SI 5). The pressures are systematically contained between 1 kbar and 1.5 kbar for both freezing rates in the entire temperature range (-15°C : -60°C). At $h$ = 500 µm (green symbols in Figure 2c,d), the primary, short-range, hydration regime, characterized by the structuring of water at the membrane surface,[38,61,62] starts to be observed at 5°C·min$^{-1}$ (Figure 2c), while it is well-identified at 10°C·min$^{-1}$ (Figure 2d), below -50°C. The pressure reached at $h$ = 500 µm remains confined below 2 kbar for both freezing rates. Closer to the cold surface, at $h$ = 100 µm, the primary hydration regime is reached for both rates below ~ -35°C. Under these conditions average pressures can reach 3 kbar at 10°C·min$^{-1}$, although the error in estimating the exact pressure becomes relevant and pressures as high as 5 kbar can actually be reached. The origin of the confidence range strongly depends on the use of four fitting strategies to fit the pressure-distance calibration curves but also, at $d_w$< 0.5 nm, on the uncertainty associated to the choice of the membrane thickness. For a thorough discussion of the origin of the uncertainty, please report to section SI 5 in the Supporting Information and to ref. [25]. The evolution of $d_{(100)}$ with temperature at 10°C·min$^{-1}$ shows that a constant value of (4.04 ± 0.01) nm (Figure 2, Figure S 2b) is reached below – 55°C, indicating that the incompressibility limit of the bilayer and its hydration layer have been reached. Reaching a plateau of the interlamellar thickness means that the lamellar probe, and consequently our model, has also reached its limits, which are set in the vicinity of -55°C at 10°C·min⁻¹ and $h$ = 100 µm (Figure 2b,d). By "reaching the limit" we mean that pressure can increase but we cannot detect it due to the incompressibility of the lamellar probe. We then set the pressure range between 2.2 kbar and 5.2 kbar as the limit allowed by the model of the lamellar probe.

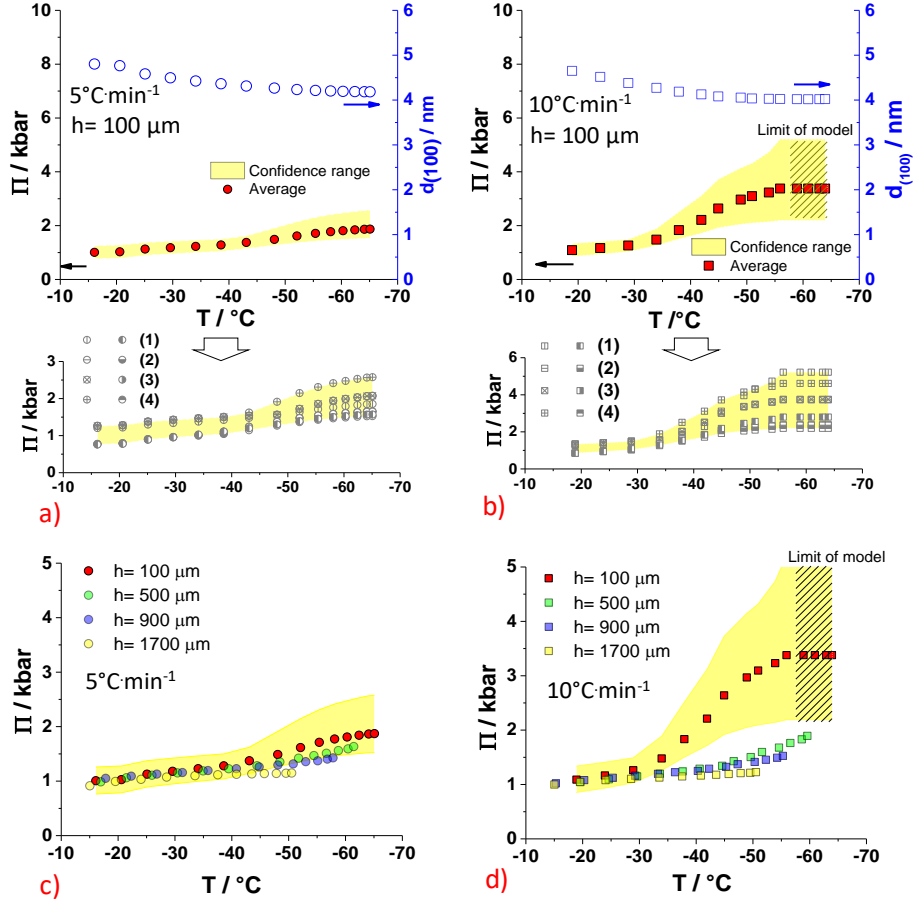

Figure 2: Pressure-temperature, $\Pi(T)$, profiles obtained for the lipid lamellar probe (GC18:0, C = 10 wt%, pH = 6.2 ± 0.3, [NaCl] = 50 mM) in a typical ice-templating experiment between -10°C and -60°C at two freezing rates and four distances, $h$, from the cooling surface (Figure S 1): a) freezing rate of 5°C·min⁻¹ (ice-front speed: 9.16 µm/s)[44], $h$ = 100 µm; b) freezing rate of 10°C·min⁻¹ (ice-front speed: 15.44 µm/s)[44], $h$= 100 µm; c) 5°C·min⁻¹ , $h$= 100 µm - 1700 µm; d) 10°C·min⁻¹ , $h$= 100 µm - 1700 µm. $\Pi(T)$ (Eq. 4) are obtained by introducing $d_w(T)$ (= $d_{(100)}(T)$-3.6 nm) in Eq. 3 according to fits (1)-(4) discussed in Ref. [25] and SI 5 in the Supporting Information. Each fit has been developed for a low- and high-ionic strength range. Full colored symbols indicate the average values calculated from all fits for a given freezing temperature and location in the ice-templating device. Yellow shades identify the upper and lower pressure limits of the fits: the bottom part of a) and b) shows how the confidence range is determined by $\Pi(T)$ plots using fits (1)-(4) at low- (intersecated symbols) and high- ionic strength (half-filled symbols) regimes (refer to Ref. [25]).

An important finding arising from the data discussed above is that, at 10 °C·min$^{-1}$ freezing rate, the pressures generated close to the cold surface (i.e., $h$ = 100 µm) are superior to the pressure attained at higher distances as depicted in Figure 3. Such difference is most likely due to supercooling of water prior to freezing in the bottom of the sample. This supercooled zone tends to crystallize suddenly, once nuclei reach a critical size, to form dendritic ice crystals over hundreds of micrometers, depending on the freezing conditions.[1] This zone is thus strongly controlled by nucleation and is characterized by rapid ice growth rates. When the supercooled zone is completely frozen, the system tends to be controlled by ice crystals' growth at fairly slow – and stable rates. These differences are often reported from the observation of the morphological differences between the bottom of ice-templated samples and their upper part.[63] Here we report that these morphological differences are accompanied by a large gap in pressure for a given temperature. At $T$= -55 °C, attained at a freezing rate of -10 °C·min$^{-1}$ , the same GC18:0 sample, measured at $h$ = 100 µm reaches 3.3 kbar whereas at $h$ = 500 µm the pressure is limited to 1.7 kbar. While these differences may seem irrelevant for they are both in the kbar-range, their implications are dramatic. When exposed to non-physiologic hydrostatic pressure, eukaryotic cells display three main responses: (i) viability, (ii) apoptosis and/or (iii) necrosis. The threshold that discriminates these responses lies in a pressure range between 1.5 and 2.5 kbar.[12] Such a threshold highlights the importance of the pressure results obtained here and, more generally, stresses the need for a strict control over the ice nucleation and growth phenomena in cryobiology and more generally in ice-templating pressure sensitive systems.

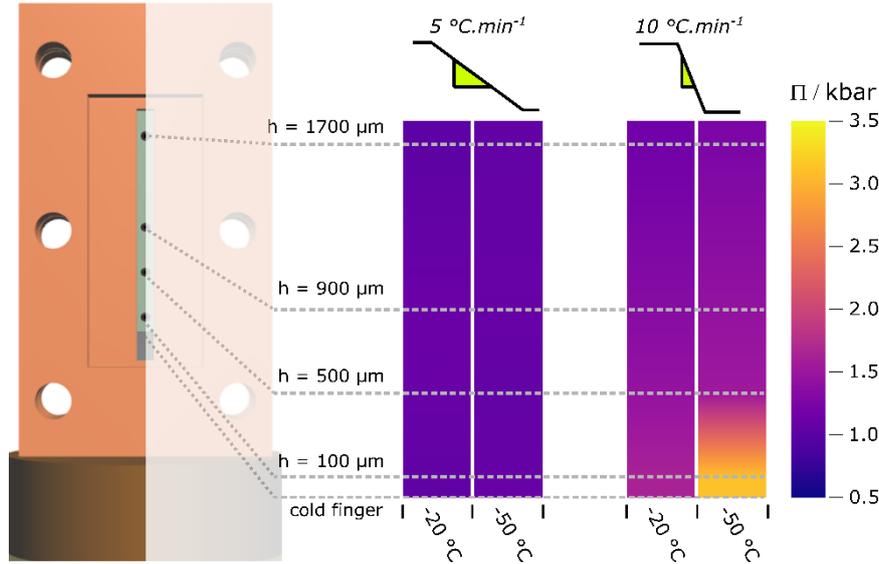

**Figure 3: Pressure produced by ice crystals growth measured during *in situ* ice-templating SAXS experiments at -20°C and -50 °C in GC18:0 samples according to the imposed freezing rate and to the distance from the cold metal surface.**

**Conclusion**

In this work we estimate within a confidence range between 10 % and 20 %, the average pressure exerted during the directional growth of ice columns when a solute composed of a glycolipid lamellar hydrogel undergoes ice-templating, also known as freeze-casting. Despite the broad interest of the material's science, soft matter and colloids, biology and medicine communities towards this technique, the values of the pressures involved during ice-templating were not known with precision at a given temperature below water crystallization. We determine the pressure in an indirect way: i) we employ temperature-resolved *in situ* SAXS experiments between +20°C and -60°C to follow the lamellar $d_{(100)}$ spacing of the lamellar hydrogel; ii) we use $^2$H solid state NMR to verify and quantify the presence of liquid-like water in the same temperature range; iii) we associate a value of osmotic pressure to a given lamellar $d_{(100)}$ distance using adiabatic dessiccation experiments, providing typical pressure-distance profiles in the same *d*-spacing range of the ice-templating experiments.

During ice-templating, we find $d_{(100)}$ spacings between 6 nm and 4 nm when the temperature drops from *T*= -15°C (crystallization starts) to -60°C. In this interval the amount of free water varies between 5 % and 1 % of the total water content and thus indicating the presence of liquid-like water, most likely localized at the interface betweenlipid and membrane. For equivalent $d_{(100)}$ spacings measured under adiabatic conditions at *T*= 25°C and relative humidity values between 10 % and 60%, we find a pressure interval contained between 1 kbar and 3 kbar. Associating temperature-resolved ice-templating experiments to the humidity-resolved adiabatic desiccation experiments, we are able to draw, for the first time in an ice-templating system, pressure-temperature profiles between $T$= -10°C and -60°C, a range of paramount importance in early stages of cryopreservation. At water crystallization ($T$= -15°C), the pressure of ice against the lamellar phase can be quantified to 1.0 ± 0.1 kbar; between $T$= -15°C and -40°C, the pressure has an average value of about 1.1 ± 0.2 kbar, while between $T$= -40°C and -60°C the pressure raises at 1.9 ± 0.4 kbar. These ranges seem to be independent of the freezing rate (here, 5°C·min$^{-1}$ and 10°C·min$^{-1}$ ), although the distance from the cold metal surface (here tested between 100 μm and 1700 μm) seems to play a major role, where the largest relative errors (up to 40%) where obtained at the closest to the finger and for fast freezing rates. The origin of such a large error is directly related to the uncertainty in the exact determination of the lipid membrane thickness; even if the uncertainty is fairly small ~0.2 nm, one expects large divergence of the short-range (water thickness < ~0.5 nm) repulsive hydration pressure when the interlamellar water layer thickness, calculated as d$_{(100)}$ minus the membrane thickness, becomes small (< ~0.5 nm) and comparable with the typical length of the interaction (below ~0.2 nm).

The data presented in this work quantify, for the first time, the pressure involved in the ice-templating process, thus serving a broad range of scientific communities and being particularly useful for the field of cryopreservation.

**Supplementary Information:** Materials and Methods,

Figure S 1

Figure S 2

Figure S 3

Figure S 4

Figure S 5

Figure S 6

Table S1 and Table S 2, Supplementary text, References for SI reference citations

**Acknowledgements:** We thank Dr. S. Roelants and Prof. Wim Soetaert (Gent University, Belgium) for producing the GC18:0 compound and Dr. E. Delbeke and Prof. C. Stevens (Gent University, Belgium) for the hydrogenation reaction. Access to the D16 beamline was financed by ILL under the proposal number 9-13-783, DOI: 10.5291/ILL-DATA.9-13-783. This work received financial support from the ESRF – The European Synchrotron, Grenoble, France, under the experiment number SC-4479. We gratefully thank Dr. Bruno Demé (ILL, Grenoble, France) for helpful critical discussion and helping us to setup the dessication experiment.

**Author contributions:** NB and FF designed, performed the experiments and wrote the manuscript. VC provided support in the neutron diffraction experiments. TZ and GBM provided assistance in the SAXS experiments. GL performed the $^2$H NMR experiments.

**Competing Interests:** All authors declare no conflicts of interests.

Supplementary Information

# Unveiling the Interstitial Pressure Between Growing Ice Crystals During Ice-Templating Using a Lipid Lamellar Probe

**Niki Baccile,[a,*] Thomas Zinn,[b] Guillaume Laurent,[a] Ghazi Ben Messaoud,[a,†] Viviana Cristiglio,[c] Francisco M. Fernandes[a,*]**

[a] Sorbonne Université, Centre National de la Recherche Scientifique, Laboratoire de Chimie de la Matière Condensée de Paris, LCMCP, F-75005 Paris, France

[†] Current address: DWI- Leibniz Institute for Interactive Materials, Forckenbeckstrasse 50, 52056 Aachen, Germany

[b] ESRF - The European Synchrotron, 71 Avenue des Martyrs, 38043 Grenoble, France

[c] Institut Laue-Langevin, 71 Avenue des Martyrs, 38042 Grenoble Cedex 9, France

**This PDF file includes:**

Materials and Methods

Figs. S1 to S6

Tables S1 and S2

Supplementary text

References for SI reference citations

**Supplementary Information Text**

**SI 1 - Materials and Methods**

*Products*. Acidic deacetylated C18:0 glucolipids (GC18:0) have been used from previously existing batch samples, the preparation and characterization ($^1$H NMR, HPLC) of which is published elsewhere.[1] Acid (HCl 37%) and base (NaOH) are purchased at Aldrich. MilliQ-quality water has been employed throughout the experimental process.

*Preparation of hydrogels*. Protocol of hydrogel preparation and characterization of the lamellar phase are reported elsewhere.[2] Shortly, GC18:0 sample is dispersed in water, followed by sonication and adjustment of pH to the desired value and ionic strength. A given amount of GC18:0 (C = 5 wt% or 10 wt%) is dispersed in a given volume of milliQ-water (generally 1 mL). The pH of the mixture is generally between 3.5 and 4.5, according to the sample concentration. The pH is then adjusted in the range 5.5 - 7.5 using 1-20 µL of NaOH 1 M (0.1 M can also be used for refinement), according to concentration and desired pH-value. The mixture is then sonicated between 15 and 20 min in a classical sonicating bath to reduce the size of the aggregated powder and until obtaining a homogenous, viscous, dispersion. To this solution, the desired volume of NaCl is added so to obtain a given total [Na$^+$] (= [NaOH] + [NaCl]) molar concentration. To keep the dilution factor negligible, we have used a 5 M concentrated solution of NaCl. The mixture is then sonicated again during 15 min to 20 min and eventually vortexed two or three times during 15 s each. The solution can then be left at rest during 15 min to 30 min. The hydrogel is a biphasic fluid containing lamellar domains and solvent. The lamellar phase is thoroughly characterized with small-angle x-ray and neutron scattering and with polarized light optical microscopy.[2,3]

*Ice-templating*. The unidirectional ice-templating/freeze-casting setup is home-built according to the literature.[4,5] The typical setup consists in a liquid nitrogen Dewar, a 40 cm copper bar (diameter: 1.5 cm), a heating element and, generally, a polypropylene tube partially inserted in the hot end of the copper bar to hold the sample prior to freezing. For this work, we build a specific ice-templating cell that could be adapted to the SAXS beamline to run *in situ* experiments.[6] The polypropylene tube has been replaced with a 2 mm flat cell, whereas the cell is supported by a plastic holder containing two face-to-face Kapton© windows. The image and scheme of the device are shown in Figure S 1. The direction of ice-templating, that is of the ice-growing front, is identified as the *Z*-axis throughout the paper. The assembly is carried out in such a manner that half of the copper bar plunges into liquid $N_2$ to create a heat sink. The temperature of the opposed extremity of the copper is controlled by the simultaneous action of the heat sink and the heating element. The heating element is controlled by a dedicated PID thermocontroller able to modulate the cooling rate (in this work, 5°C·min$^{-1}$ and 10°C·min$^{-1}$). A temperature sensor (K thermocouple) is located at the bottom of the cell, close to the tip of the copper bar. In a typical experiment, 1 mL or 2 mL of the hydrogel is poured inside the sample holder, in direct contact with the copper surface.

*Small Angle X-ray Scattering (SAXS)*: SAXS experiments have been performed on the ID02 beamline at the ESRF – The European Synchrotron in Grenoble, France. The experiments have been done at a photon energy of 17.0 keV ($\lambda$ = 0.7 Å) for two sample-to-detector distances 2 m and 8 m, respectively. Calibration of the *q*-range is done using silver behenate ($d_{(001)}$ = 58.38 Å). The scattering data is recorded with a Rayonix MX-170HS CCD detector. The raw data are normalized and integrated azimuthally using standard procedures. The one-dimenional scattered intensity $I(q)$ is given in dimensionless units subtracted by the water background. The magnitude of the scattering vector $|\mathbf{q}| = q$ is given by $q = 4\pi/\lambda \sin(\theta)$ with the scattering angle $2\theta$. Typical acquisition times are in the order of 100 ms, which we considered enough to obtain a good signal-to-noise ratio where no beam damage is observed. One spectrum per temperature value is recorded.

The SAXS patterns are recorded at a step of 5°C in the entire temperature range at different positions in the cell: for each temperature, the signal is collected at four positions, $h$, (Z-axis) simultaneously, namely, $h$ = 100 (the closest to the bar), 500, 900, 1700 μm from the top of the copper bar (Figure S 1b), respectively. The movement of the stage along Z axis is controlled by an automated motorized sample stage available at the beamline. Both the acquisition time and the stage displacement are fast enough (< 1 s, including signal acquisition and displacement for the four positions) with respect to the cooling rate, which is 0.17 °C/s for the fastest rate. In view of these considerations, one can consider that the measurement can be considered instantaneous (the sample is in the same physical state for all positions at a given measurement time) for all positions.

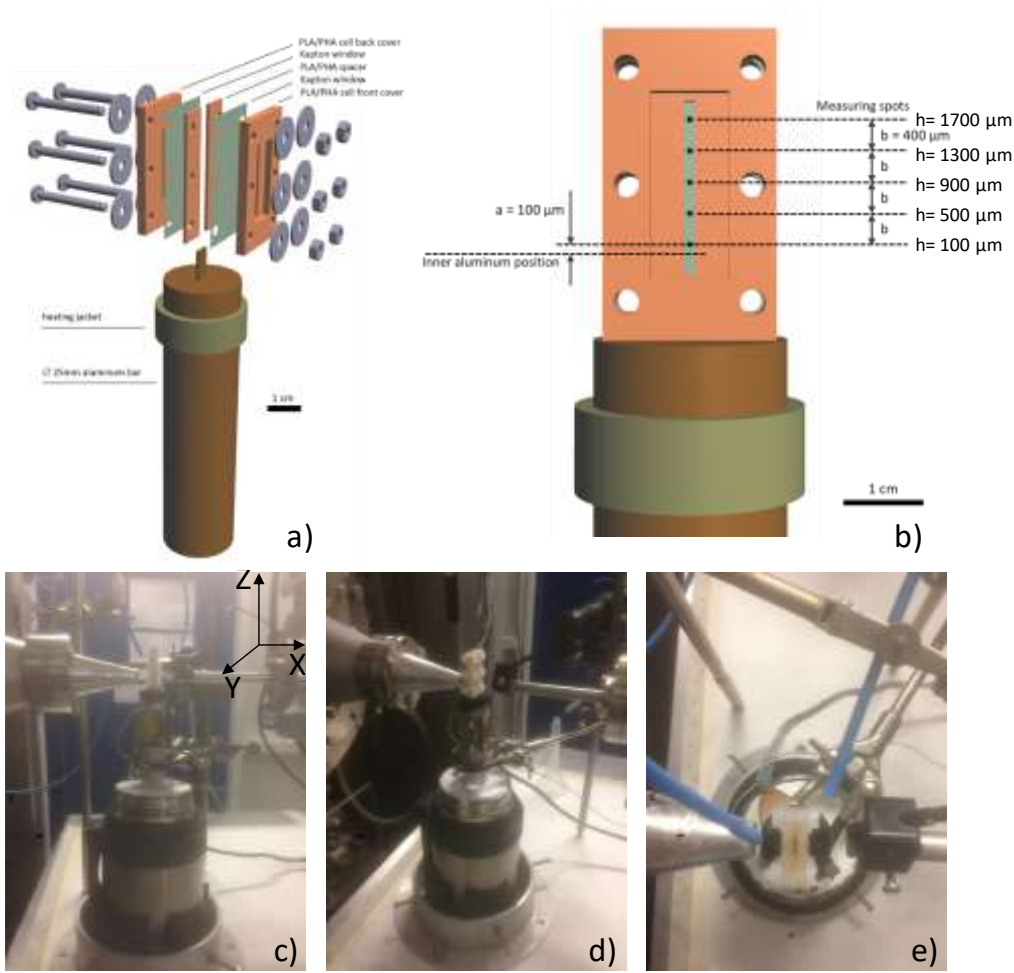

**Figure S 1: a)** Ice-templating setup used in the *in situ* SAXS experiments. The bottom part of the aluminum bar (not shown) is kept in liquid nitrogen. The heating jacket is controlled by a dedicated PID and the temperature sensor (K thermocouple) is located at the top of the aluminum bar. **b)** Detail of the in situ freeze-casting cell used in synchrotron experiments. The cell is assembled from 3D printed PVA/PHA parts and kapton tape, assembled by 6 M6 nylon screws and knobs. The measuring position spots within the cell are indicated by black dots. **c-e)** Side, front and top views of the freeze-casting cell coupled to the ID02 beamline at the ESRF. The liquid nitrogen Dewar is located at the bottom of the cell while the X-ray source is on the right-hand side. The blue pipes in (e) carry a constant dry air flux in the front and bottom of the cell to avoid condensation and crystallization of moisture.

*Humidity chamber experiments*. Two 1 wt% GC18:0 solutions are prepared in $D_2O$ at pH = 6.2 and at [NaCl] = 16 mM and 100 mM. The solutions are deposited on two separate 5 cm x 2 cm silicon wafers by simple drop cast (volume dropped: 500 µL). To enhance homogeneous spreading of the solution onto the substrate, we have used a horizontal support levelled with a 2D spirit level. The silicon substrates are let dry in an oven at 40°C until a homogeneous coating is obtained. The samples are then introduced within a humidity chamber (Figure S 4a), provided at the beamline, and set under vacuum at T= 25°C. The temperature of the $D_2O$ water bath below the sample is modified to set the chamber at the desired relative humidity, *RH%*, value. The humidity chamber is conceived to provide values of RH% with an error of ±0.01% RH. Technical details of the humidity chamber can be found in ref.[7] The sample at [NaCl]= 16 mM is let equilibrating at 98 *RH%* before studying, where relative humidity is lowered. The sample at [NaCl]= 100 mM sample is let equilibrating at 10 *RH%* and humidity is increased.

*Neutron diffraction*: neutron diffraction experiments are carried out as described in ref.[8] on the D16 instrument at the Institut Laue-Langevin (ILL; Grenoble, France), using a wavelength $\lambda$= 4.5 Å ($\Delta\lambda/\lambda$= 0.01) and a sample-to-detector distance of 900 mm.[9] The focusing option provided by the vertically focusing graphite monochromator is used to maximize the incident neutron flux at the sample. The intensity of the diffracted beam is recorded by the millimeter-resolution large-area neutron detector (MILAND) $^3$He position-sensitive detector, which consists of 320 × 320 *xy* channels with a resolution of 1 × 1 mm$^2$. The samples are held vertically in a dedicated temperature-controlled humidity chamber and aligned on a manual 4-axis goniometer head (Huber, Rimsting, Germany) embedded in the humidity chamber. The chamber is mounted on the sample rotation stage, where the lipid multilayer stacks are scanned by rocking the wafers horizontally. Diffraction data are collected at a detector angle 2 $\theta$ of 12˚, by scanning the sample angle $\omega$ in the range -1 to 8˚, with a step of 0.05˚. Data analysis is performed using the ILL in-house LAMP software (www.ill.eu/instruments-support/computing-for-science/cs-software/all-software/lamp).[10] The classical *I* vs *2θ* profile for each *RH%* is obtained by summarizing each integrated 2D image measured at a given value of $\omega$. The lamellar spacing $d_{(100)}$ is obtained by a fitting the (100) peak position with a Gaussian profile. Intensities on the detector surface are corrected for solid angle and pixel efficiency by normalization to the flat incoherent signal of a 1 mm water cell.

The sample temperature in the chamber is maintained at 25°C during the measurements, and the humidity is varied by changing the temperature of the liquid reservoir generating the water vapor from 10°C to 24°C, leading to relative humidities ranging from to 10% to 98%. Each sample is investigated by increasing the humidity step by step without opening the chamber at any time during the humidity scan. After each change in relative humidity, the sample is equilibrated between 30 min to 2 h, where equilibration is followed through the evolution of the (100) diffraction peak position over time. Equilibration time is followed (by collecting $\omega$-2θ scans) until the diffraction peak position reach a plateau. After equilibration, the rocking curve ($\omega$ scan between -1° and 8° with 0.05°) is recorded.

*²H solid state Nuclear Magnetic Resonance (NMR)*: ²H solid-state NMR experiments are performed on an AVANCE III 300WB Bruker spectrometer ($B_0$= 7.0 T) with a VTN dual HX probe-head, and a 4 mm zirconia rotor. Temperature is regulated with Topspin 3.6.1 software, a BCU-Xtreme accessory, and a bearing pressure of 3 bars. To ensure static operation (no spinning at magic angle is imposed), a thin layer of Teflon is added on cap wings. After 5 minutes at +20°C, temperature is ramped down at 10°C·min$^{-1}$. Below -55°C, probe thermal inertia limited cooling speed. Finally, temperature is kept 5 minutes at -60 °C. The probe is tuned at -5°C. For each acquisition, 20 transients are collected with a π/2- π/2 solid-state echo, pulses of 3.45 µs, a half-echo delay of 10 µs, a spectral with of 1 MHz, and a recycling delay of 3 s, giving 1 minute per spectrum. Temporal signals are left shifted by 6 points, to avoid first order phasing, and exponential apodization of 1 kHz is applied before Fourier transform. Spectra are fitted with dmfit software,[11] with a Lorentzian component for the narrow peak and a quadrupolar static shape for the broad peak, with a quadrupolar constant of 203 kHz and an asymmetry parameter of 0.

## SI 2 - Distance-temperature profiles from ice-templating *in situ* SAXS

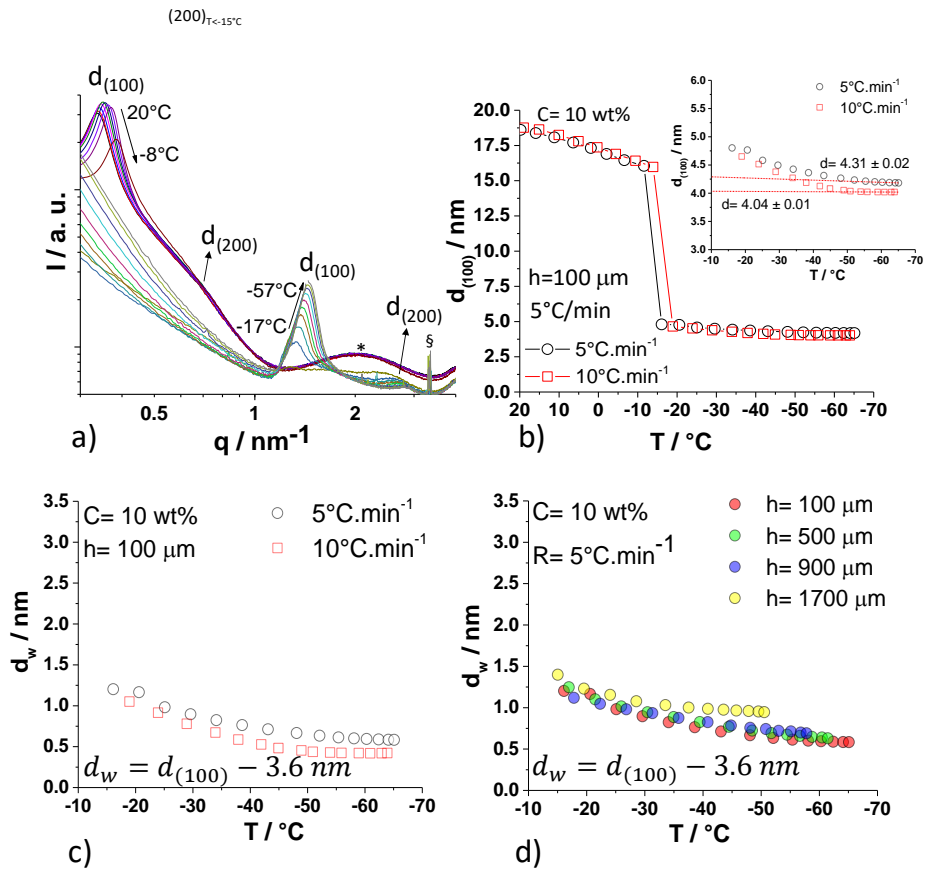

**Figure S 2: a)** Typical SAXS profiles (background not subtracted) recorded on a GC18:0 lamellar hydrogel (C = 10 wt%, pH = 6.2 ± 0.3, [NaCl] = 50 mM) in the ice-templating device shown in Figure S 1 at freezing rate of 5°C·min$^{-1}$ and at the position of $h$ = 900 μm in the ice-templating device. SAXS profiles are given as a function of temperature. * indicates the oscillation of the membrane form factor[1,2,12]; § indicates artifacts. **b)** Evolution of the $d_{(100)}$ distance with temperature. **c-d)** Evolution of the interlamellar thickness, $d_w$, in the temperature region below -10°C.

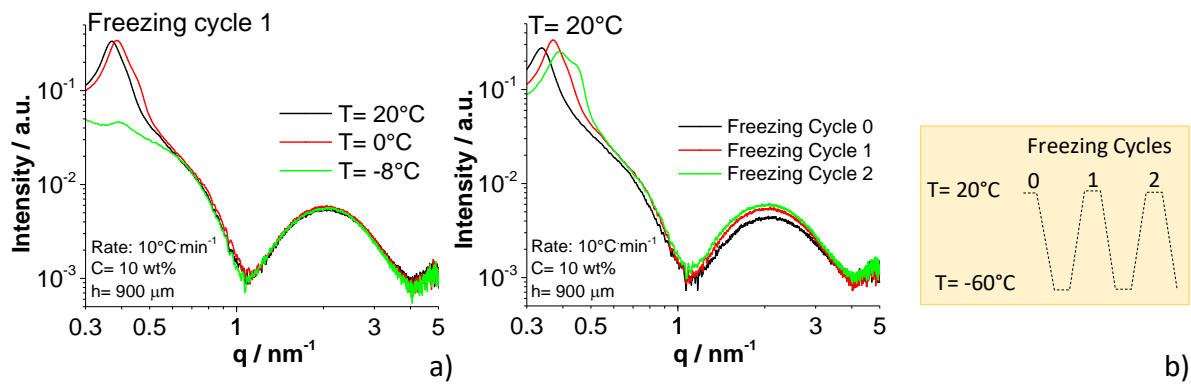

**Figure S 3:** Background-subtracted SAXS profiles recorded on a GC18:0 lamellar hydrogel (C = 10 wt%, pH = 6.2 ± 0.3, [NaCl] = 50 mM) in the ice-templating device shown in Figure S 1 at freezing rate of

**10°C·min⁻¹ and at the position of the ice-templating device, $h$ = 900 μm. In panel a) the SAXS profiles are shown during a given freezing cycle from $T$= +20°C to the value of $T$= -8°C, the last one available before ice crystallization. Panel b) shows three SAXS profiles recorded at T= +20°C before (black curve, freezing cycle 0) and after two freezing cycles (red and green curves). The profile of the freezing cycles (freezing rate: 10°C·min⁻¹) is also given in b).**

Figure S 3a shows the typical SAXS profiles of the GC18:0 lamellar hydrogel recorded from room temperature to $T$= -8°C, this temperature corresponding to the last value before ice crystallization and before the abrupt variation of $d_{(100)}$ from $d$ ~16 nm to $d$ ~5 nm, as shown in Figure S 2a,b. Figure S 3b shows three SAXS profiles recorded at $T$= +20°C on the same sample before and after two complete freezing cycles from room temperature to $T$= -60°C, as indicated in the profile of the freezing cycles in Figure S 3b. All SAXS profiles in Figure S 3 are typical of the GC18:0 lamellar hydrogel, largely discussed elsewhere: the $d_{(100)}$ of the lamellar order at low-$q$ coexists with a broad oscillation characteristics of the interdigitated GC18:0 lipid layer at $q$> 0.5 nm⁻¹.[1–3,12]

As a whole, Figure S 3 shows that the SAXS profiles of the lipid lamellar hydrogel are all superimposable above $q$> 0.5 nm⁻¹, the $q$-region that is characteristics of the thickness of the lipid membrane, found to be 3.6 nm for this system.[1,12] In particular, Figure S 3a shows that the minimum and amplitude of the oscillation at $q$> 0.5 nm⁻¹, associated to the form factor (including thickness and electron density) of a bilayer, are exactly the same. This evidence reinforces our hypothesis according to which the thickness of the lipid membrane is independent of temperature, at least down to T= -8°C, and also reasonably below this value. Unfortunately, we cannot provide reliable SAXS profiles below T= -8°C due to the strong scattering of ice, which masks the lipid signal, as shown in Figure S 2a for $T$≤ -17°C. Figure S 3b shows, as an additional support, that the SAXS profile at $q$> 0.5 nm⁻¹ of the lipid lamellar hydrogel is also practically unchanged after two complete freezing cycles from T= +20°C to T= -60°C. This fact supports the idea that the interdigitated GC18:0 lipid membrane is not disrupted nor perturbated by the ice-templating process.

# SI 3 - Pressure-distance profiles from neutron diffraction osmotic pressure experiments in the humidity chamber

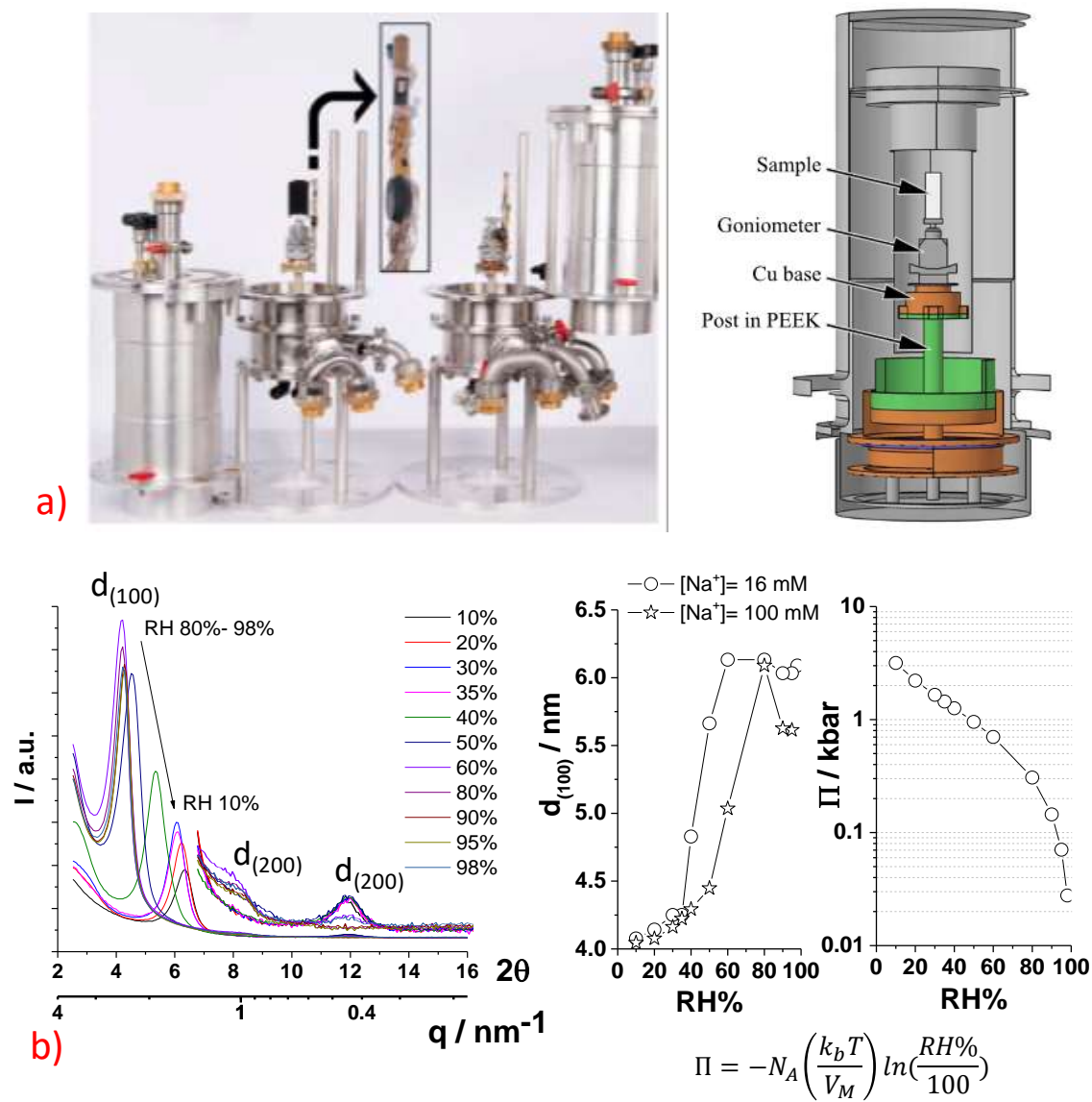

**Figure S 4: a) Humidity chamber used on the D16 beamline at ILL and used to perform adiabatic desiccation experiments.[7] b) Evolution of the diffraction patterns as a function of relative humidity, *RH*%, measured on a GC18:0 solution (bulk data: C= 1 wt%, pH 6.2 ± 0.3, [NaCl]= 16 mM) drop-cast on a silicon (111) substrate. Evolution of the $d_{(100)}$-spacing with *RH*%, and plot of the corresponding $\Pi(RH\%)$ relationship with $N_A$ being the Avogadro's number, $K_b$ the Boltzmann constant, *T* the temperature in Kelvin degrees and $V_m$ the water molar volume.**

Pressure-distance experiments are performed at T= 25°C by controlling the relative humidity under adiabatic conditions[7] on GC18:0 samples prepared at (bulk data) C= 1 wt%, pH 6.2 ± 0.3 and [NaCl]= 16 and 100 mM. In synthesis, intensity against scattering angle and treated data, $d_{(100)}(RH\%)$ and $\Pi(RH\%)$, are shown in Figure S 4b. In Ref. 13, we have extensively treated these data in order to obtain analytical expressions of the pressure-distance, $\Pi(d_w)$, profiles, which can be summarized in Eq. S 1, with $\Pi_{VdW}$ being the attractive Van der Waals interaction, and $\Pi_{Hyd1}$ and $\Pi_{Hyd2}$ being the primary and secondary repulsive hydration interactions, of which the general expression is given in Eq. S 2, where $\Pi_H$ is the hydration pressure, $\lambda_H$ the decay length and $d_w$ the interlamellar water thickness.

$$\Pi(d_w) = \Pi_{VdW} + \Pi_{Hyd1} + \Pi_{Hyd2} \qquad \text{Eq. S 1}$$

$$\Pi(d_w)_{Hyd} = \Pi_H e^{-\frac{d_w(RH\%)}{\lambda_H}} \qquad \text{Eq. S 2}$$

We have employed four fitting strategies to analyze the pressure-distance data and to obtain analytical expressions of $\Pi(d_w)$, as extensively described and commented in Ref. 13. The values of the strength and decay length of the primary and secondary hydration interactions are summarized for each fit in Table S 1.

**Table S 1 : Values of the hydration pressure ($\Pi_H$) and decay lengths ($\lambda_H$) in the primary and secondary hydration regimes. The general expression of $\Pi(d_w)_{Hyd}$ is given in Eq. S 2. Data are obtained from the fit of osmotic stress experiments, $\Pi(d_w)$, applying fits (1)-(4) and recorded for the GC18:0 lamellar hydrogel prepared under low- (16 mM) and high-salt (100 mM) conditions. The fits and the full approach are described in Ref. 13.**

| Fit N° | [Na$^+$] / mM | $\Pi_{H1}$/kbar | $\lambda_{H1}$/nm | $\Pi_{H2}$/kbar | $\lambda_{H2}$/nm |
|---|---|---|---|---|---|
| (1) | 16 | 1.26·10$^3$ | 0.07 ± 20% | 1.81 | 2.98 ± 20% |
| | 100 | 37.5 | 0.13 ± 10% | 1.61 | 1.59 ± 10% |
| (2) | 16 | 17.0 ± 40% | 0.28 ± 20% | 2.04 ± 15% | 2.53 ± 20% |
| | 100 | 5.94 ± 7% | 0.45 ± 10% | 1.80 ± 12% | 1.43 ± 10% |
| (3) | 16 | 17.3 ± 40% | 0.28 ± 20% | 2.04 ± 15% | 2.50 ± 20% |
| | 100 | 6.05 ± 7% | 0.45 ± 10% | 1.81 ± 2% | 1.42 ± 10% |
| (4) | 16 | 7.93·10$^6$ | 0.28 ± 20% | 8.47 | 2.52 ± 20% |
| | 100 | 1.66·10$^4$ | 0.45 ± 10% | 8.47 | 1.42 ± 10% |

# SI 4: $^2$H-NMR experiments during freezing under non-equilibrium conditions

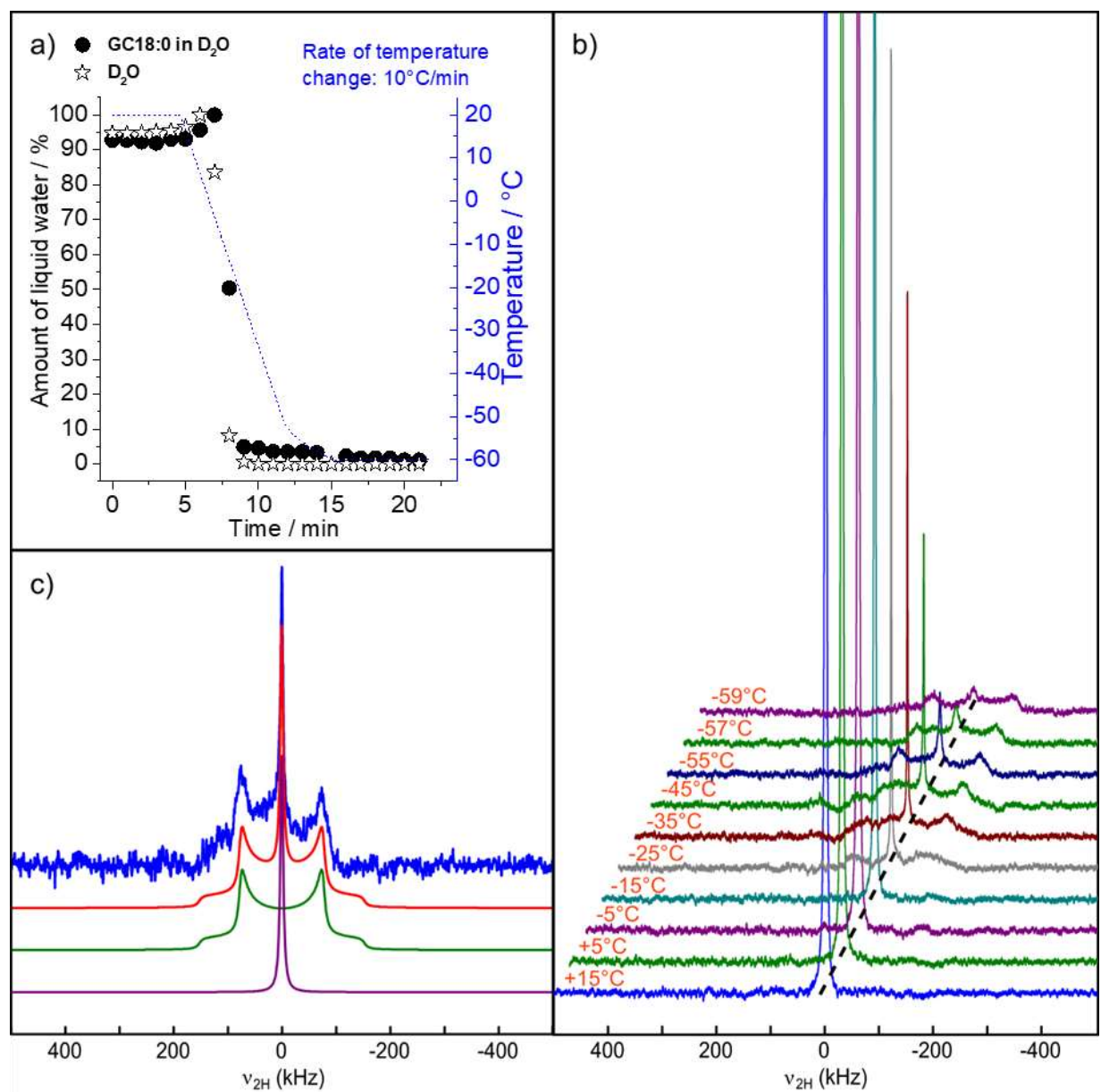

**Figure S 5:** $^2$H solid-state NMR experiments performed during freezing from +20°C to -60°C at a rate of 10°C·min$^{-1}$ for a GC18:0 lamellar hydrogel (C = 5 wt%, pH = 6.2 ± 0.3, [NaCl] = 300 mM) prepared in 100% D$_2$O. Temperature is controlled directly in the NMR probe using the BCU-Xtreme temperature unit. The sample is placed in a typical zirconia 4 mm-rotor and the temperature is measured outside the rotor. a) Evolution of the amount of liquid water in the lamellar hydrogel (circles, left) and in pure D$_2$O (stars, left) and corresponding temperature profile (dashed line, right). The amount represents the normalized integrals of the resonance at 0 kHz (Lorentzian component corresponding to liquid-like water) of the $^2$H-NMR spectra. A slight increase of the measured amount of liquid water was observed before freezing. This could be explained by a faster longitudinal T1 relaxation rate when decreasing temperature and mobility, favoring quantification. b) Selected $^2$H-NMR spectra of the lamellar hydrogel and relative averaged temperature corresponding to the data in a); a dashed line guides the eye to follow the central resonance at **0 kHz. c)** Typical fit (red) of a lamellar hydrogel spectrum recorded below the freezing point (-55°C, in blue) is constituted of two components, a Lorentzian one at 0 kHz, corresponding to liquid water (in purple), and a broad quadrupolar signal, corresponding to solid water (in green). Typically, an apodisation of the Lorentzian component of 8 ppm, a quadrupolar coupling of 203 kHz and an asymmetry parameter of 0 are used to fit the quadrupolar component.

**Table S 2.** Liquid water peak relative intensity of $^2$H solid-state NMR experiments performed at equilibrium between +20°C and -60°C: 20 min thermalization is employed between two consecutive temperature steps of 10°C. Data corresponds to GC18:0 lamellar hydrogel (C = 5 wt%, pH = 6.2 ± 0.3, [NaCl] = 300 mM) prepared in 100% D$_2$O and pure D$_2$O. The asterisk, *, indicates that these values are approximate and should be taken as an upper estimate due to the poor signal-to-noise ratio in these experiments.

| T / °C | Normalized intensity % (GC18:0) | Normalized intensity % (D$_2$O) |
|---|---|---|
| 20 | 87.7 | 92.5 |
| 10 | 91.6 | 96.3 |
| 0 | 95.1 | 100.0 |
| -10 | 100.0 | 0.5* |
| -20 | 6.6 | 0.5* |
| -30 | 5.0 | 0.5* |
| -40 | 2.3 | 0.0 |
| -50 | 0.8 | 0.0 |
| -60 | 0.0 | 0.0 |

**SI 5 - Determining $\Pi(T)$ profiles by associating distance-temperature, $d_w(T)$, experiments from ice-templating and pressure-distance, $\Pi(d_w)$, experiments from humidity chamber**

Ice-templating *in situ* SAXS experiments performed on the GC18:0 lamellar hydrogel (C = 10 wt%, pH = 6.2 ± 0.3, [NaCl] = 50 mM) shown in Figure S 2 generate a set of $d_{(100)}(T)$ profiles, which can be plot in terms of $d_w(T)$, that is interlamellar thickness as a function of temperature, for an interdigitated layer thickness of 3.6 nm. $d_{(100)}(T)$ (or $d_w(T)$) are generated for two freezing rates, 5°C·min$^{-1}$ and 10°C·min$^{-1}$, and four heights in the ice-templating device, $h$= 100, 500, 900 and 1700 µm (refer to Figure S 1).

For each experimental $d_{(100)}(T)$ plot we calculate $d_w(T)$ as indicated in Eq. S 3; then, from Eq. S 1, we generate a general equation $\Pi(T)$ (Eq. S 4) by substituting the water thickness with $d_w(T)$. Each term of Eq. S 4 will also depend on temperature, as shown in Eq. S 5, Eq. S 6 and Eq. S 7 for, respectively, $\Pi(T)_{VdW}$, $\Pi(T)_{Hyd1}$ and $\Pi(T)_{Hyd2}$. For a given fitting strategy developed in Ref. 13 and NaCl concentration provided in Table S 1, we then generate the corresponding $\Pi(T)$ profile according to the following:

- For fit (1), we employ equation Eq. S 4 - Eq. S 7. Parameters for $\Pi_{VdW}$ (Eq. S 4) are the Hamaker constant, $H$ = 5.1·10$^{-21}$ J, thickness of the hydrophilic layer of the lipid membrane, $T_h$= 1.4 nm, and thickness of the hydrophobic layer of the lipid membrane, $L$= 0.8 nm. Parameters for $\Pi_{Hyd1}$ and $\Pi_{Hyd2}$ are given in fit (1) line in Table S 1. *Note on the value of the Hamaker constant, $H$*: The Hamaker constant can be precisely calculated at room temperature,[14,15] but one must recall that $H$ depends on temperature and should be calculated for T< 0°C. The calculation of $H$ for a lipid bilayer can be found in ref. [14,15] and it is the sum of zero frequency ($\frac{3}{4}k_bT\left(\frac{\epsilon_1-\epsilon_2}{\epsilon_1+\epsilon_2}\right)^2$) and frequency-dependent ($\frac{3}{16\sqrt{2}}h\nu_e\frac{(n_1^2-n_2^2)^2}{(n_1^2-n_2^2)^{3/2}}$) contributions, with $h$ being the Planck constant in J·s, $\nu_e$ the absorption frequency, $\epsilon_1, \epsilon_2$ the dieletric constants and $n_1, n_2$ the refractive indexes of lipid (subscript 1) and medium (subscript 2). At room temperature, one can safely take $\epsilon_1$= 2, $\epsilon_2$= 80, $n_1$= 1.46, $n_2$= 1.33 for $\nu_e$ of 60 THz. Upon freezing, one must use the dielectric constant ($\epsilon_2$= 90)[16] and refractive index ($n_2$= 1.35) for ice, while variations for the lipid phase are practically negligible. The final effect on $H$ is in the order of 10-15%, a fact which has no practical influence in the final calculation of $\Pi_{VdW}$. For this reason, we use $H = 5.1 \cdot 10^{-21}$ J throughout this work.

- For fit (2), we employ equations Eq. S 6 and Eq. S 7. Parameters for $\Pi_{Hyd1}$ and $\Pi_{Hyd2}$ are given in fit (2) line in Table S 1.

- For fit (3), we employ equations Eq. S 6 and Eq. S 7. Parameters for $\Pi_{Hyd1}$ and $\Pi_{Hyd2}$ are given in fit (3) line in Table S 1, while parameters for $\Pi_{VdW}$ (Eq. S 5) are $H = 5.1 \cdot 10^{-21}$ J; $T_h = 1.4$ nm; $L = 0.8$ nm.

- For fit (4), we apply Eq. S 3 and the pressure is traced directly against $d_{(100)}(T)$: $\Pi(d_{(100)})$. We employ equations Eq. S 6 and Eq. S 7. Parameters for $\Pi_{Hyd1}$ and $\Pi_{Hyd2}$ are given in fit (4) line in Table S 1.

$$d_w[T] = d_{(100)}[T] - 3.6 \; nm; \;\; T < -10 \; °C \hspace{2cm} \text{Eq. S 3}$$

$$\Pi(T) = \Pi(d_w[T]) = \Pi(T)_{VdW} + \Pi(d_w[T])_{Hyd1} + \Pi(d_w[T])_{Hyd2} \hspace{1cm} \text{Eq. S 4}$$

$$\Pi(T)_{VdW} = \frac{H(T)}{6\pi}\left(\frac{1}{d_w^3} - \frac{2}{(d_w(T) + 2T_h + L)^3} + \frac{1}{(d_w(T) + 2(T_h + L))^3}\right) \hspace{0.5cm} \text{Eq. S 5}$$

$$\Pi(T)_{Hyd1} = \Pi_{H1} e^{-\frac{d_w(T)}{\lambda_{H1}}} \; ; \; d_w < 0.74 \pm 0.11 \; nm \hspace{2cm} \text{Eq. S 6}$$

$$\Pi(T)_{Hyd2} = \Pi_{H2} e^{-\frac{d_w(T)}{\lambda_{H2}}} \; ; \; d_w > 0.74 \pm 0.11 \; nm \hspace{2cm} \text{Eq. S 7}$$

*General note on the fitting strategy*: The drawback of fits (1)-(3) is the plot of the pressure against the water thickness, $d_w$, being calculated from the thickness of the interdigitated GC18:0 membrane, only estimated here at room temperature but not below water crystallization. Unfortunately, precise measurement of the bilayer structural parameters ($T_h$ and $L$) after ice crystallization should be performed but it is a complex task because the strong scattering of ice masks the signal of the membrane. One must then formulate the hypothesis that $T_h$ and $L$ (and, consequently the total thickness, $2T_h+L$) do not vary much at temperatures well below zero degree. Although quite a strong assumption, this hypothesis is not outrageous for three reasons: the lipid membrane has low volumetric compressibility, as assumed by Wolfe;[17,18] $T_h$ and $L$ are already measured for the GC18:0 membrane about 15°C below its $T_m$,[2,12] that is in a gel rigid configuration; the oscillation above $q= 0.5$ nm$^{-1}$, typical of the bilayer form factor, stays unchanged from room temperature until $T= -8$°C, just before crystallization (Figure S 3a). As commented by Wolfe, lipid bilayers in the gel state are less prone to be damaged, or undergo structural changes, than membranes in the fluid state.[17] In any case, to avoid the uncertainty of knowing the thickness of the membrane at temperature below the freezing point, we have employed fit (4), where pressure data are plotted in a log-lin representation against the interlamellar distance, $d_{(100)}$, and not the water thickness. Structural parameters $\Pi_{H1}$, $\Pi_{H2}$, $\lambda_{H1}$, $\lambda_{H2}$ are simply extracted from a double linear fit.[13] One should note that in fit (4) we neglect the Van der Waals contribution, as we do in fit (2), and that in fit (4) only the slopes, from which $\lambda_{H1}$, $\lambda_{H2}$ are obtained, are significant, while the pressure values at the intercept, $\Pi_{H1}$, $\Pi_{H2}$, are not.

Finally, for a given freezing rate and height in the ice-templating device, we obtain a set of eight $\Pi(T)$ plots (four fits, two salt conditions), of which the average value (red circles) is plot in Figure 2 in the main text and Figure S 6. At the same time, the eight $\Pi(T)$ plots are systematically used to generate a confidence band (yellow), as also illustrated in Figure 2 in the main text and Figure S 6.

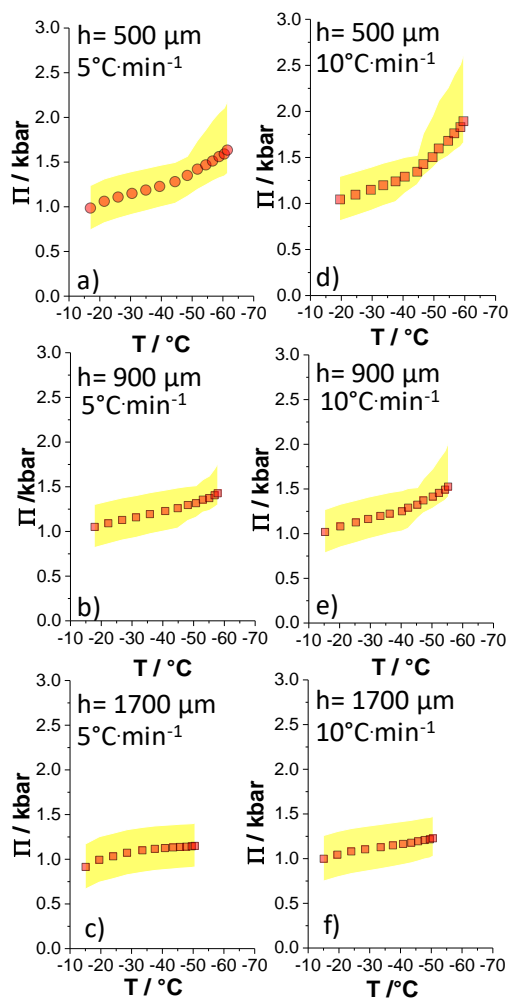

**Figure S 6:** Pressure-temperature, $\Pi(T)$, profiles obtained for the lipid lamellar probe (GC18:0, C = 10 wt%, pH = 6.2 ± 0.3, [NaCl] = 50 mM) in a typical ice-templating experiment between -10°C and -60°C at two freezing rates and four distances, $h$, from the cooling surface (Figure S 1): the conditions are indicated on each panel. $\Pi(T)$ (Eq. S 4) are obtained by introducing $d_w(T)$ (Eq. S 3) in Eq. 3 (main text, $\Pi(d_w) = \Pi_{VdW} + \Pi_{Hyd1} + \Pi_{Hyd2}$) according to fits (1)-(4) discussed in Ref. 13. Each fit has been developed for a low- and high-salt range. Full symbols indicate the average values calculated from all fits for a given freezing temperature and location in the ice-templating device. Yellow shades identify the upper and lower pressure limits of the fits. Fits are described in Ref. 13.